# Heart rate variability code:
# Does it exist and can we hack it?

Martin G. Frasch
Dept. of Obstetrics and Gynecology
University of Washington School of Medicine
Seattle, WA, USA
mfrasch@uw.edu   |   0000-0003-3159-6321

*Abstract*—Code is generally defined as a system of signals or symbols for communication. I synthesize the experimental evidence for the presence and utility of such communication in the heart rate variability (HRV): HRV contains signatures of information flow between the organs and of response to physiological or pathophysiological stimuli as signatures of states (or syndromes). HRV exhibits features of time structure, phase space structure, specificity with respect to (organ) target and pathophysiological syndromes as well as universality with respect to species independence. Together these features form a spatiotemporal structure, a phase space, that can be conceived of as a manifold of a yet to be fully understood dynamic complexity. The objective of this article is to deliver a proof-of-concept that physiological evidence supports the existence of HRV code whereby the process-specific subsets of HRV measures indirectly map the phase space traversal reflecting the specific information contained in the code required for the body to regulate the physiological responses to those processes. I review the following physiological examples of HRV code reflected in specific changes to HRV properties across the signal-analytical domains: the fetal systemic inflammatory response, the organ-specific inflammatory responses (brain's and gut's), chronic hypoxia and intrinsic (heart's own) HRV (iHRV), allostatic load (physiological stress due to surgery) and vagotomy (bilateral cervical denervation). I propose future studies to test these observations in more depth and refer the interested reader to the referenced publications for a detailed study of the HRV measures involved. The presented framework promises more specific HRV biomarkers of health and disease.

## I. Introduction

Heart rate variability (HRV) derived from ultrasound-based or electrocardiogram (ECG)-based sensors represents the key source of predictive information about fetal well-being, development and early detection of physiological abnormalities. Psychophysiological models of HRV have been proposed and used intensely in research.[2,3] Many mathematical features of HRV have been reported and for some, a connection to parts of physiology has been established. However, an overarching concept of how to understand the rich mathematical structure of HRV as a specific reflection of the brain-body communication has been missing. Here, I attempt to address this gap by introducing and defining specifically the notion of HRV code.[1] My approach is summarized in Figure 1.

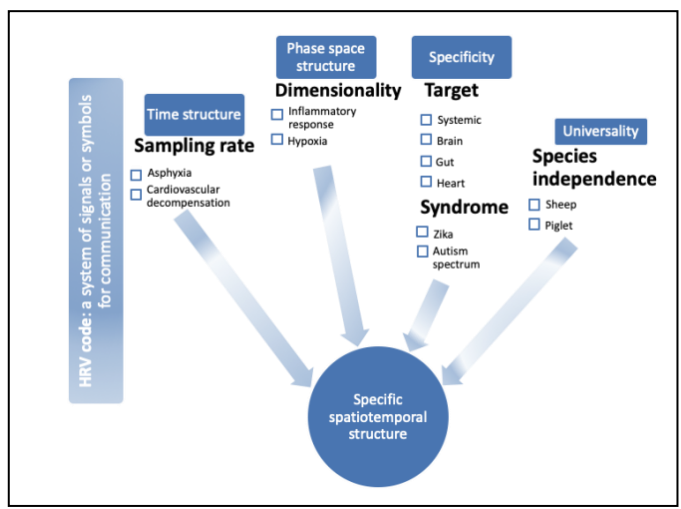

Fig. 1. **Heart rate variability (HRV) code.** HRV exhibits features of time structure, phase space structure, specificity with respect to (organ) target and pathophysiological syndromes as well as universality with respect to species independence. These properties represent the necessary features of an HRV code of brain-body communication. Under each feature, physiological examples are given and discussed in the article.

## II. Time structure: sampling rate

### A. Evidence

In studies on chronically instrumented fetal sheep mimicking human labor [4] and in human studies with HRV recordings during labor [5], we found that the sampling rate of the signal used to construct the beat-to-beat variability time series, i.e., the underlying electrocardiogram (ECG, 1000 Hz) or CTG (cardiotocography, 4 Hz) signals that render HRV, impact the precision of predicting the onset of acidemia, a fetal condition unique to labor and that sometimes is associated with unfavorable outcomes. Similarly, the early detection of fetal cardiovascular decompensation, a sentinel of incipient brain injury due to reduced cerebral blood flow, is impeded by the lower sampling rate of the underlying fetal ECG signal.[6] As putative underlying mechanisms, we proposed the reduced precision in capturing the coding of vagally mediated signaling in response to acidemia or Bezold-Jarisch reflex, a vagally mediated adaptive cardiac depressor reflex preserving injured myocardium.[7]–[9]

Generically for any data capture per Nyquist-Shannon sampling theorem, if the information is indeed encoded in the HRV signal, it must be sampled correctly for true representation to contain complete predictive power.[10]

In several studies using animal models and human cohorts, we made observations validating this notion. Studies in pregnant sheep demonstrate the potential to detect fetal acidemia and predict cardiovascular decompensation during labor using high-precision ECG recording of fetal heart rate (FHR).[4,6,9] A novel bioinformatics approach to fetal HRV (fHRV) derived from maternal abdominal ECG during labor predicted well acid-base status at birth.[5]

These insights do not rule out the potential for predictive utility of the ultrasound-derived FHR, while pointing to possible intrinsic limitations of different sensor technologies with regard to their ability to adequately capture salient physiological information as well as the risk of capturing physiological artifacts at lower sampling rates.[4,6,11]

*B. Significance*

The impact of the duration and sampling rate of ECG on the quality and representativeness of the ECG-derived HRV measures has been well-documented in the adult HRV studies, but has remained largely neglected in the fetal HRV literature, in part due to the sheer inaccessibility of high-quality fetal ECG technology in the clinical or research settings. With this in mind, the notion of time structure in HRV may appear not novel from the perspective of the "adult" HRV studies, but it does represent the cutting edge in fetal monitoring. Moreover, within the integrative framework of HRV code (Fig. 1) the property of time structure exemplified but not limited to the effects of the sampling rate represents a novel model component, not described as such before.

III. PHASE SPACE STRUCTURE: DIMENSIONALITY OF HRV

Here I use the notion of dimensionality as it pertains to characterizing a system's phase space.[12] I will review evidence that HRV shows a complex phase space structure, i.e., including time dimension. Assuming that HRV activity represents an integrated set of properties of the physiological system, the information in the HRV is organized such that, with enough observations, HRV properties across the signal-analytical domains capture the entire phase space the system can traverse. Consequently, if the HRV code is adequately resolved, a set of deterministic equations or machine learning models can predict the behavior of the underlying system.

An important and enigmatic contributor to HRV properties is the vagus nerve activity. The neuroanatomical distribution of the vagus nerve goes well beyond the well-known respiratory and cardiovascular controls as well as the more recently discovered immunological control. The vagus nerve connects the brain with the body's each and every organ, in particular, thymus, paraganglia (distributed sensor system in thorax and abdomen), liver, gastrointestinal tract, uterus, pancreatic islets, chemoreceptors sensing nutrients and related compounds (glucose, amino acids, fatty acids, and neuropeptides), mechanosensors (touch, tension, serosal), temperature sensors, osmosensors, nociceptors.[13] Despite this detailed understanding of neuroanatomy, with the exception of nutrient [14, 15] and glucosensing [16–18] the functional implications of this highly distributed innervation have remained largely unknown.

*A. Evidence*

Physiological predictions are possible under the assumption of HRV phase space structure and can shed light on the complex vagal nerve's contributions to HRV. Let us focus on HRV behavior in the following physiological examples:

- The systemic inflammatory response [19], [20];
- Organ inflammatory response: brain and gut [21];
- Chronic hypoxia and intrinsic HRV (iHRV) [22, 23];
- Allostatic load: physiological stress due to surgery [24];
- Vagotomy (bilateral cervical denervation) [24].

The cholinergic anti-inflammatory pathway (CAP) signals via the vagus nerve's afferent branch to surveil the body's inflammatory milieu and relay this information to the brain.[25, 26] Increased inflammation results in increased vagal neural outflow dampening the inflammation. The afferent vagus nerve projects to nuclei in the brainstem, hypothalamus, amygdala, insular and cingulate cortices, but the precise brain's centers involved in processing afferent vagal information remain largely unknown. The indirect data reported so far suggest the existence of a neuroimmunological homunculus.[13], [27–30]

In contrast to the afferent and cerebral processing networks, the role of the efferent branch of the vagus nerve in the fine-grained control of inflammatory milieu under physiological and pathophysiological conditions has been studied more extensively. Underscoring its homeokinetic role, the research on the efferent branch yielded detailed insights into the anti-inflammatory, but not immunosuppressive, effects of the vagus nerve signaling on spleen's macrophages.[26,31] In parallel, evidence in animal models and human studies has accrued that such neuroimmunological signaling is also reflected in the specific changes of HRV properties.[19–21], [32–36]

*B. Significance*

The pattern of changes in HRV traverses a not well-understood phase space that can be, at least in part, characterized by complementary HRV measures from different signal-analytical domains (cf. Table S1 in [24]). Such putative regions of the phase space capture specifically systemic or organ-specific, gut or brain, inflammatory responses to lipopolysaccharide (LPS), because they reflect the underlying HRV code of the brain-body communication.[19]–[21]

In addition to predicting inflammation, another subset of HRV measures characterizes the iHRV properties imprinted by chronic exposure to hypoxia.[22] A complex landscape of HRV measures characterizes the chronic effects of stress imposed on the body due to surgery and that landscape shifts subtly but distinctly when the surgery is conducted with the initial bilateral cervical removal of the vagus nerve.[24]

IV. TARGET AND SYNDROME SPECIFICITY

*A. Evidence*

The subsets of the HRV measures reflecting gut or brain inflammation referred to above are specific to those targets. That

I gratefully acknowledge the funding from: CHU Ste-Justine Research Center, Molly Towell Perinatal Research Foundation, CIHR, FRQS, QTNPR (CIHR), NeuroDevNet, MITACS, UW Global Innovation Fund, Dept. of OBGYN, University of Washington, Seattle.

is, for gut and brain, no HRV measure correlated to several markers of inflammation at the same time.

These target-specific observations in gut and brain inflammation suggested the possibility of identifying further target specific HRV signatures. While the efforts to identify HRV code represent an indirect attempt to hack brain-body communication, the direct data obtained from vagus nerve electroneurograms (VENG) studies is also mounting to suggest the existence of a vagus code.[16,37,38] It is plausible to expect a reflection of such vagus code in HRV properties that vagus activity is known to modulate and more studies are needed to connect specific VENG features with multi-dimensional HRV properties.[24,39]

If we are to postulate code specificity, target specificity must be complemented by another kind of unique encoding property: specificity to identifying a complex physiological or pathophysiological behavior or a syndrome. To that claim, I will mention once more the finding of iHRV as well as the recent findings of HRV changes in the toddlers exposed to Zika during gestation, and eight years old children with autism spectrum disorder (ASD), conduct disorder and depression.

Sheep fetuses exposed to chronic hypoxia in vivo and their hearts mounted in Langendorff apparatus ex vivo, showed changes in cardiac activity: the beat-to-beat variability of the isolated hearts from hypoxic fetuses demonstrated the existence of a signature of such exposure.[22] Interestingly, the mathematical properties of the HRV measures comprising this signature were also found in an independently conducted prospective study of human toddlers born without overt symptoms to mothers exposed to the Zika virus during pregnancy.[40] Zika virus infection can cause chronic hypoxia via its effects on placental function, so the iHRV signature of fetal hypoxia is relevant for the changes in HRV due to Zika.[41] Together, these findings suggest that chronic hypoxia alters the cardiac pacemaker cellular activity and synchronization impacting the properties of the emerging beat-to-beat variability. Another possible impact is on the normal developmental program of the pacemaker cell genes such that the resulting beat-to-beat variability is imprinted by the exposure carrying de facto a memory of hypoxia.[22]

*B. Significance*

Further studies are needed to better understand the impact of various insults on the developmental and functional properties of individual and synchronized cardiac pacemaker cells yielding the heartbeat automaticity and intrinsic variability. This will help develop a stronger causal understanding of how the various environmental exposures impact the HRV properties with regard to HRV code target specificity.

In the cohort of 68 children, machine learning models built from features representing HRV measures from five signal-analytical domains computed from five minute ECG recordings identified children from healthy, ASD, conduct disorder and depression cohorts with the area under the receiver operating curve of more than 0.82.[42] The fundamental role of vagus nerve in ASD has been subject of ongoing research, notably in the polyvagal theory.[2,43,44] This finding highlights the potential of the HRV code approach to capture complex traits or syndromes.

It has been proposed that ASD represents a process defined by a set of developmental immunometabolic constraints that may explain much of its endophenotypic features.[45] HRV has been shown to exhibit dynamic patterns with metabolic challenges.[46–50] Does HRV reflect the metabolic status and metabolic optimization as do genomic adaptations?[51–53] Future studies should systematically investigate the relationship between HRV code and immunometabolism, especially in ASD.

Can this information now be used to predict trajectories toward these conditions at an earlier age so as to help commence more timely interventions? If the HRV code persists and has memory properties, this should be a reasonable expectation and tested in future studies.

## V. Universality

*A. Evidence*

Another anticipated aspect of an HRV code should be its universality, i.e., patterns of communication reflected in HRV code would apply across species, at least to an extent that they represent some phylogenetically preserved or functionally parsimonious solutions. That is, it seems likely that complex systems would converge onto similar if not identical solutions to a problem such as encoding multi-organ communication, especially when relying on a similarly implemented substrate, such as the vagus nerve. We tested this in a study on newborn piglets exposed to a sublethal dose of LPS.[54] We applied the same HRV inflammatory index known to track systemic inflammatory response to LPS in fetal sheep[19,20] to the newborn piglet's HRV. We found that the HRV inflammatory index tracked the inflammatory response in this species correctly following the temporal inflammatory profile of the measured cytokines. These findings encourage the notion of HRV code universality, at least in the application to tracking inflammation.

*B. Significance*

An important aspect of the study design, when investigating the universality of HRV code, is the known between-subject variability of HRV measures.[55] This can be compensated by a within-subject design with repetitive measurements creating within-subject temporal profiles coordinated between subjects in terms of diurnal data collection, as for example, done in the above-mentioned fetal sheep and piglet studies or in the review by Laborde et al.

Further studies are required to test these observations more extensively, in different species, and relate them to the corresponding changes in VENG properties.

## VI. Discussion

We have observed that HRV exhibits a rich spatiotemporal structure if captured at adequate temporal resolution. That is, HRV properties are phase-space- and, in particular, time-resolution-dependent. These insights inform practical considerations of data acquisition and analysis such as the sampling rate, choice of HRV measures and the machine learning models to predict the outcome or classify subjects based on the outcome. I mentioned some applications of machine learning to both prediction [6,9] and classification [42] problems using HRV data accounting for these considerations.

The subsets of HRV measures relate to specific physiological responses and hence are akin to code with regard to target specificity. This has implications for HRV data interpretation and biomarker discovery using HRV technology.

A fascinating question needs to be answered in future studies: after more than fifty years of studying HRV behavior under various conditions, can we bring this knowledge base together with the emerging field of bioelectronic medicine and its many promising therapeutic approaches to create a closed-loop system where HRV or VENG code hacking are used to sense the behavior of the system of interest and the vagus nerve stimulation or other bioelectronic tools are used to modify its phase space dynamics?

Reproducing the above observations under different physiological conditions and in different species will help address the issue of HRV code robustness. Does the HRV code benefit from stochastic resonance?[56] Can this property be measured in HRV, in vivo or in silico?

At last, while the above considerations are driven by the pragmatic pursuit of HRV code applications for predictive purposes of time series, there remains the big question of what HRV code really represents physiologically. We can take inspiration for conceptualizing HRV code from the brain's default-mode network concept.[57] Does the HRV code exhibit spontaneous intrinsic dynamics that are altered by external stimuli? I believe the answer to this can already be stated in the affirmative. Progress has been made in defining the cardiac origins of HRV, arising from stochastic fluctuations of cardiac pacemaker cells' ion channel activity, their networked synchronization and modulation of this synchronization dynamics by the milieu intérieur and external neural and humoral inputs.[58] Future studies will address the questions whether we can formulate the exact gestalt of such intrinsic dynamics driven by homeostasis and inter-organ communication on one hand, and define the activation patterns within the HRV code in response to external stimuli on the other hand.

ACKNOWLEDGMENT

In this paper, I integrate findings of multiple studies and am grateful to collaborators and their amazing teams: Dr. Andrew Seely (Ottawa Health Research Institute), Clinical Sciences of the Faculté médecine vétérinaire (Université de Montréal), Dr. Enrico Ferrazzi (Dept. of Obstetrics and Gynecology, University of Milan), Dr. Steven Wang (Dept. of Mathematics and Statistics, York University), Dr. Hau-Tieng Wu (Depts. of Mathematics and Statistics, Duke University), Dr. Dino Giussani (Dept. of Physiology, Development and Neuroscience, University of Cambridge), and Dr. Schuler Faccini (Dept. de Genetica, Universidade Federal do Rio Grande do Sul).

COMPETING INTERESTS

MGF holds an ECG-related patent (WO2018160890A1) and a fetal EEG patent (US9215999).